\documentclass[prx,showkeys,showpacs,footnoteinbib,twocolumn,unsortedaddress,floatfix]{revtex4-1}

\usepackage{todonotes}

\usepackage{times}
\usepackage{amsmath}
\usepackage{amsfonts}
\usepackage{amssymb}
\usepackage{epsfig}
\usepackage{siunitx}
\usepackage{gensymb}

\usepackage{float}
\usepackage{ifthen}
\usepackage{xspace}
\usepackage{relsize}

\usepackage{algorithm}
\usepackage{algorithmic}

\begin{document}

\title{Comparing fractional quantum Hall Laughlin and Jain topological orders with the anyon collider}
\author{ M. Ruelle$^{1}$, E. Frigerio$^{1}$,   J.-M. Berroir$^{1 }$,   B.
Pla\c{c}ais$^{1 }$,   J. Rech$^{2}$, A. Cavanna$^{3}$, U. Gennser$^{3}$, Y. Jin$^{3}$, and G. F\`{e}ve$^{1 \ast}$ \\
\normalsize{$^{1}$ Laboratoire de Physique de l\textquoteright Ecole normale sup\'erieure, ENS, Universit\'e
PSL, CNRS, Sorbonne Universit\'e, Universit\'e Paris Cit\'e, F-75005 Paris, France}\\
\normalsize{$^{2}$ Aix Marseille Univ, Universit\'{e} de Toulon, CNRS, CPT, Marseille, France. }\\
\normalsize{$^{3}$ Centre de Nanosciences et de Nanotechnologies (C2N), CNRS, Universit\'{e} Paris-Saclay, 91120 Palaiseau, France. }\\
\normalsize{$^\ast$ To whom correspondence should be addressed; E-mail:  gwendal.feve@ens.fr.}\\}

\begin{abstract}
Anyon collision experiments have recently demonstrated the ability to discriminate between fermionic and anyonic statistics. However, only one type of anyons associated with the simple  Laughlin state at filling factor $\nu=1/3$ has been probed so far. It is now important to establish anyon collisions as quantitative probes of fractional statistics for more complex topological orders, with the ability to distinguish between different species of anyons with different statistics. In this work, we use the anyon collider to compare the Laughlin $\nu=1/3$ state, which is used as the reference state, with the more complex Jain state at $\nu=2/5$, where low energy excitations are carried by two co-propagating edge channels. We demonstrate that anyons generated on the outer channel of the $\nu=2/5$ state (with a fractional charge $e^*=e/3$) have a similar behavior compared to $\nu=1/3$, showing the robustness of anyon collision signals for anyons of the same type. In contrast, anyons emitted on the inner channel of $\nu=2/5$ (with a fractional charge $e^*=e/5$) exhibit a reduced degree of bunching compared to the $\nu=1/3$ case, demonstrating the ability of the anyon collider to discriminate not only between anyons and fermions, but also between different species of anyons associated with different topological orders of the bulk. Our experimental results for the inner channel of $\nu=2/5$ also point towards an influence of interchannel interactions in anyon collision experiments when several co-propagating edge channels are present. A quantitative understanding of these effects will be important for extensions of anyon collisions to non-abelian topological orders, where several charged and neutral modes propagate at the edge.
\end{abstract}
\maketitle

\section{Introduction}

Two-dimensional systems can host quasiparticles with quantum statistics intermediate between fermions and bosons \cite{Leinaas77,Wilczek82}. As the phase $\varphi$ accumulated by the wavefunction when exchanging the relative positions of two particles can take any value ($0 \leq \varphi \leq \pi$), these quasiparticles have been named anyons \cite{Wilczek82b}. The fractional value of $\varphi/\pi$ has important consequences when one performs a braiding operation, which consists in moving one particle around another one, thereby accumulating the phase $2\varphi$. In the case of fermions ($\varphi=\pi$) or bosons ($\varphi=0$), the accumulated braiding phase is trivial, with $e^{i2\varphi} = 1$. By contrast, anyons keep a memory of braiding operations as $e^{i2\varphi} \neq 1$. The stability of the braiding phase with local deformations of the anyon trajectories is at the origin of topologically protected quantum computing  using non-abelian anyons \cite{Kitaev03}.

Soon after the prediction of their existence, it was realized that anyons are the elementary excitations of fractional quantum Hall (FQH) states \cite{Halperin84, Arovas84} (for a review see \cite{Stern08}).  Different FQH states, reached by varying the filling factor $\nu$, are characterized by different topological orders \cite{Wen95} associated with different species of anyons. The Laughlin states \cite{Laughlin83}, for which $\nu=1/m$,   have the simplest topological order characterized by the single number $m$ that sets the Hall conductance $G/G_0=1/m$ (where $G_0=e^2/h$ is the conductance quantum), the fractional charge of the anyons $e^*/e=1/m$, and their fractional statistics $\varphi/\pi=1/m$. The simple topological order also implies that the edge structure is simple, with a single channel of conductance $G_0/m$ at the edge of the FQH state. The Jain sequence \cite{Jain89}, with $\nu=p/(2mp \pm 1)$ (such as $\nu=2/3$ or $\nu=2/5$), has a more complex topological order characterized by a matrix \cite{Wen95}. It implies that the fractional charge of anyons and their fractional statistics are characterized by different numbers. It also implies that the edge structure is more complex, with several co- or counter-propagating channels at the edge of the sample. Finally, the $\nu=5/2$ state \cite{Willett87} is predicted to have a non-abelian topological order \cite{Moore91}, as confirmed by thermal conductance measurements \cite{Banerjee18}.

If the existence of anyons was confirmed more than 20 years ago by the measurement of their fractional charge \cite{dePicciotto97,Saminadayar97}, their fractional statistics were only confirmed recently by two experiments  \cite{Bartolomei20,Nakamura20} (for a review of experiments probing fractional charge and fractional statistics, see \cite{Feldman21}). Ref.\cite{Nakamura20} investigated manifestations of fractional statistics using single-particle Fabry-Perot interferometry \cite{Chamon97}, whereas Ref.\cite{Bartolomei20} investigated these manifestations using two-particle Hanbury-Brown and Twiss interferometry \cite{Safi01,Vishveshwara03,Kim05,Vishveshwara10,Campagnano12,Campagnano13} in the geometry of the anyon collider \cite{Rosenow16}. These two experiments have focused so far on only one type of anyons in the simplest case of the Laughlin state at filling factor $\nu=1/3$. It is now important to establish these new experimental tools as quantitative probes of fractional statistics, with the ability to distinguish between different species of anyons for different topological orders.

In this work, we use the anyon collider to investigate and compare different species of abelian anyons. Filling factor $\nu=1/3$ is used as a reference state. Because of the simple nature of its topological order and of its edge structure, it is used for extensive tests of quantum models of anyon collisions \cite{Rosenow16,Morel22,Lee22}. We compare these measurements with the more complex topological order of the $\nu=2/5$ state, described by two co-propagating edge channels. Collision experiments performed on the outer channel of $\nu=2/5$ provide very similar results compared to the $\nu=1/3$ state. This is not surprising, as the outer channel of $\nu=2/5$ has similar properties to $\nu=1/3$ (same conductance $G_0/3$ and the same anyon fractional charge $e^*=e/3$ \cite{Griffiths00,Kapfer19}). Collision experiments performed on the inner channel of $\nu=2/5$ provide clear quantitative differences with the $\nu=1/3$ case, as expected since the nature of anyons is different, with a fractional charge $e^*=e/5$  \cite{Reznikov99,Griffiths00,Kapfer19}. Our results demonstrate the ability of the anyon collider to provide quantitative distinct signatures between different species of anyons with different statistics. They also suggest that the quantitative description of anyon collisions at $\nu=2/5$ is more complex, and that other mechanisms need to be taken into account, such as interactions between neighboring edges \cite{Levkivskyi08}, which are known to be important in the context of collision experiments \cite{Wahl14,Ferraro14,Marguerite16,Idrisov22}.

\section{The anyon collider}

\subsection{Device and principle of the experiment}

\begin{figure}[h]
\includegraphics[width=1
\columnwidth,keepaspectratio]{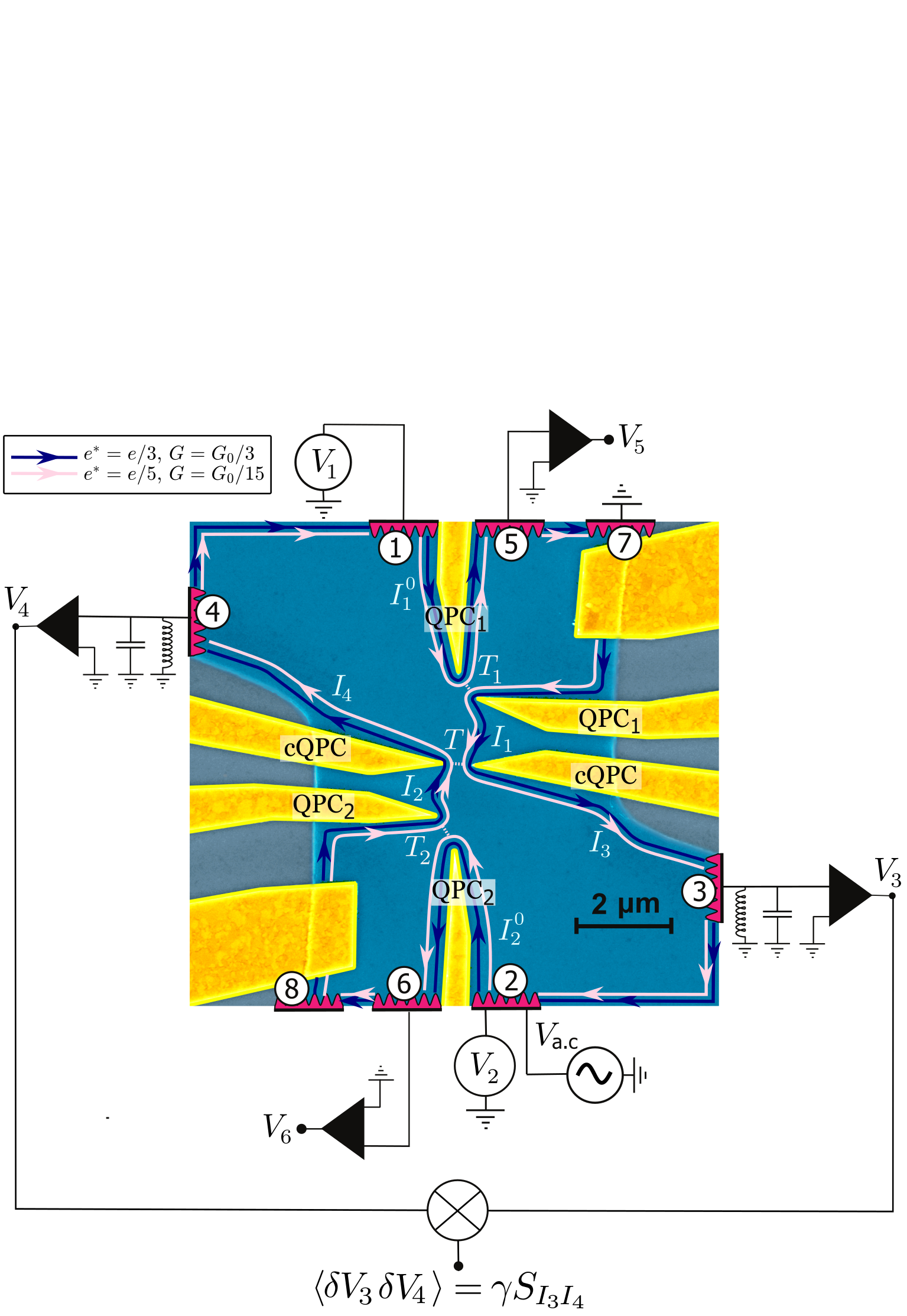}
\caption{\label{fig1} \textbf{Anyon collider device.} \\
Colored scanning electron microscope picture of the anyon collider device. The two-dimensional electron gas is represented in blue. At filling factor $\nu=2/5$, charge propagates along two co-propagating edge channels. The inner channel (pink lines) has a conductance $G_{\text{in}}=G_0/15$ (where $G_0=e^2/h$ is the conductance quantum) and carries anyons of charge $e^*=e/5$. The outer channel (blue lines) has a conductance $G_{\text{out}}=G_0/3$ and carries anyons of charge $e^*=e/3$. The dc voltages $V_1$ and $V_2$ generate the noiseless currents $I_1^0$ and $I_2^0$ towards QPC1 and QPC2 tuned in the weak backscattering regime ($T_1$, $T_2 \ll 1$). The anyon currents $I_1$ and $I_2$ propagate towards cQPC used as a beam-splitter of backscattering transmission $T$. $T_1$ (resp. $T_2$) is extracted from the measurements of the small backscattered a.c. currents $\delta I_5$ (resp. $\delta I_6$) into contact 5 when a small a.c. current $\delta I_7$ (resp. $\delta I_8)$ is injected towards QPC1 (resp. QPC2), with $T_1= \delta I_5/ \delta I_7$ (resp. $T_2= \delta I_6 / \delta I_8$).  $T$ is extracted from the measurement of the small ac current $\delta I_3$ resulting from the small current $\delta I_2^0$ injected by the ac voltage $\delta V_2$: $T=\frac{\delta I_3}{T_2 \delta I_2^0}$. Finally the current correlations $S_{34}$ between the currents at the output of the splitter are converted to voltage cross-correlations on $RLC$ tank circuits where $R=R_K/\nu$ is the Hall resistance. The conversion factor $\gamma$ is calibrated from thermal noise measurements (see Ref.[\onlinecite{Bartolomei20}]). }
\end{figure}

The anyon collider device is based on a two-dimensional electron gas at the interface of a GaAs/AlGaAs heterostructure with charge density $n_s=1.1\times10^{15}$ m$^{-2}$ and mobility $\mu=1.4\times 10^6$ cm$^2$.V$^{-1}$s$^{-1}$. Fig.\ref{fig1} shows a scanning electron microscope picture of the device. The central quantum point contact, cQPC, is used as the beamsplitter in the collision experiment. The measurement of the cross-correlations $S_{34}$ of the current fluctuations at outputs 3 and 4 of the collider provides information on the tendency of particles to bunch together or to exclude each other. Triggered single anyon sources have been theoretically proposed \cite{Ferraro15}, but they have not yet been experimentally realized. Instead, we use two QPCs, QPC1 and QPC2, tuned in the weak backscattering regime, as random Poissonian anyon sources \cite{Martin05}. Applying the d.c. voltage $V_1$ (resp. $V_2$) to ohmic contacts 1 (resp. 2), the noiseless current $I_{1}^0$ (resp. $I_{2}^0$) flows towards QPC1 (resp. QPC2) where its backscattering with probability $T_1$ (resp. $T_2$) \cite{Note1} leads to the random generation of the anyon current $I_1$ (resp. $I_2$) in the weak backscattering limit ($T_1$, $T_2 \ll 1$). The anyon currents $I_1$ and $I_2$ then propagate towards cQPC where the collision occurs.

As theoretically predicted in Ref.\cite{Rosenow16} and experimentally observed in \cite{Bartolomei20}, the current cross-correlations in an anyon collision are proportional to the total anyon input current $I_+=I_1+I_2$ via a Fano factor $P$ defined as $P=S_{34}/\big[2e^* T(1-T)I_+\big]$, where  $T$ is the backscattering transmission of cQPC. $T$ is defined as the small variation of the backscattered current $\delta I_3$ resulting from a small anyon current $\delta I_2$ at input 2.  It is measured by applying a small a.c. voltage $\delta V_2$ at ohmic contact 2 (see Fig.\ref{fig1}), leading to a small a.c. modulation of the injected current $\delta I_2^0$, with  $T=\delta I_3/\delta I_2=\delta I_3/(T_2 \delta I_2^0)$.

The anyon collider can be tuned in two different regimes. The balanced collider corresponds to equal anyon currents at the inputs of cQPC, $I_1=I_2$. It is obtained by tuning QPC1 and QPC2 at identical emission probabilities, $T_1=T_2=T_S$, such that the current difference between inputs vanishes, $I_-=I_1-I_2=0$. This configuration provides immediate qualitative differences between the behaviors of fermions and anyons in a collision. Fermionic antibunching results in a suppression of the cross-correlations in the balanced case, $P(I_-=0)=0$. On the contrary, as discussed in Ref.\cite{Rosenow16}, anyons are allowed to form packets of charge in a given output. This results in negative current cross-correlations, leading to negative values of $P$. Another interesting configuration is the unbalanced collider, which corresponds to $I_- \neq 0$. The level of imbalance between the two sources can be tuned  by the ratio $I_-/I_+$. $I_-=I_+$ corresponds to switching off one source. This configuration provides both distinct experimental signatures between fermions and anyons, and between different species of anyons with different statistics, with the possibility to compare quantitatively experimental signals with quantum models of anyon collisions.

\subsection{Elements of theory}

 As discussed in \cite{Morel22,Lee22}, the mechanisms governing anyon bunching have a different nature than the ones responsible for fermion antibunching. Introducing the tunneling Hamiltonian at cQPC, $H_T=A+ A^{\dag}$, where $A$ describes the creation of an anyon in output 3 and of its hole counterpart in output 4 (with tunneling amplitude $\zeta$), the dominant contribution to the out of equilibrium temporal correlations of the tunneling processes at cQPC can be computed \cite{Rosenow16,Morel22,Lee22} in the long time limit $t\gg h/(e^*V)$:
  \begin{eqnarray}
\langle A^{\dag}(0) A(t) \rangle_{\text{neq}} & =& e^{-N_1(t) (1-e^{-2i\pi \lambda } )  } \times e^{-N_2(t) (1-e^{2i\pi \lambda})}  \nonumber\\
& \times & \langle A^{\dag}(0) A(t) \rangle_{\text{eq}} + \text{subleading terms}  \label{tunnel_noneq}
\end{eqnarray}
$\langle A^{\dag}(0) A(t) \rangle_{\text{eq}}$ are the temporal correlations of tunneling processes at equilibrium, $N_1(t)$ (resp. $N_2(t)$) is the average number of anyons randomly emitted by QPC1 (resp. QPC2) in time $t$. Finally $e^{2i\pi \lambda}$ is the braiding factor for anyons at the edge. The parameter $\lambda$ is introduced in Eq.(\ref{tunnel_noneq}) since the braiding phase $2\pi \lambda$ for anyons at the edge may differ from the braiding phase $2\varphi$ for anyons in the bulk. This may occur when the edge structure is complex, due to the topological order enforcing the presence of several edge channels or due to edge reconstruction mechanisms. Coulomb interaction between edge channels may then lead to charge fractionalization mechanisms \cite{Berg09,Kamata14,Inoue14,Freulon15}. The resulting fractionalized charges may have different mutual fractional statistics \cite{Lee20}, resulting in a modified value of the parameter $\lambda$ at the output of the interedge interaction region \cite{Levkivskyi16,Idrisov22}. The Laughlin case $\nu=1/m$ is the simplest regime where a single channel is present and interaction mechanisms may be neglected. In this reference situation, one expects  $\lambda = \varphi/\pi=e^*/e=1/m$. The case of  $\nu=2/5$ is more complex, due to the presence of two co-propagating edge channels, and there are no predictions for the value of $\lambda$ in this case yet.

As discussed in Refs.\cite{Morel22,Lee22}, the presence of the braiding factors in Eq.(\ref{tunnel_noneq}) can be interpreted as resulting from braiding mechanisms, occurring in the time-domain, between anyons generated by the input QPCs and anyons transferred at cQPC. As a result, $P(I_-=0)$ can be expressed as a function of $\lambda$ and of the exponent for anyon tunneling $\delta$, which governs the long time decay of the correlations in the fractional state\cite{Wen91}:
\begin{eqnarray}
P(I_-=0) & = & 1-\frac{\tan{(\pi \lambda)}}{\tan{(\pi \delta)}}\frac{1}{1-2\delta} \label{P}
\end{eqnarray}
As discussed above for the parameter $\lambda$, the tunneling exponent $\delta$ is also related to the topological order in the bulk in the case of a simple edge structure. In particular, one expects $\delta=1/m$ in the Laughlin case, but $\delta$ may also be affected by edge reconstruction mechanisms \cite{Rosenow02}. Laughlin states can thus be seen as reference states for comparisons with quantum models of anyon collisions, with $\lambda=\delta=\varphi/\pi=e^*/e=1/m$. As already mentioned, the edge structure for $\nu=2/5$, with two co-propagating edge channels, is more complex. Possible values for $\lambda$ and $\delta$ at $\nu=2/5$ are discussed in sections \ref{Bal_2_5} and \ref{Unbal_2_5}, where collision experiments at $\nu=2/5$ are compared to the $\nu=1/3$ case on the same sample.

The unbalanced case, $I_- \neq 0$, also offers a striking way to distinguish between fermions and anyons and between different species of anyons. In the anyon case, as seen in Eq.(\ref{tunnel_noneq}), braiding mechanisms occur in different directions for anyons emitted by QPC1 (with a braiding phase $-2\pi\lambda$) and anyons emitted by QPC2 (with a braiding phase $+2\pi\lambda$). This means that the contribution of both sources is not additive and one expects interferences between both sources tuned by the ratio $I_-/I_+$. In particular, one expects $|P|$ to decrease when $I_-/I_+$ decreases since braiding mechanisms occur in opposite directions for the two anyon sources (as observed in \cite{Bartolomei20}). Qualitatively, this evolution of $P$ with $I_-/I_+$ is a signature of braiding mechanisms, as opposed, for example, to the recently observed Andreev scattering at a QPC \cite{Glidic22}, which occurs in a different limit where the input QPCs, QPC1 and QPC2, do not scatter the same fractional charge as the cQPC. In the case of Andreev processes, no interferences between the two sources are expected and the output cross-correlations should therefore not depend on the imbalance ratio $I_-/I_+$. Quantitatively, the exact dependence of $P(I_-/I_+$) with $I_-/I_+$  is directly related to  the values of the parameters $\lambda$ and $\delta$ and as such, provides a way to discriminate between different species of anyons.

In the electron case, one also expects a dependence of $P$ with $I_-/I_+$, as fermion antibunching is suppressed when one source is switched off (e.g. source 2), resulting in a restoration of the negative cross-correlations for $I_-=I_+$. However, in the electron case, the negative cross-correlations have a different origin. As braiding mechanisms are absent ($e^{2i\pi \lambda} = 1$ for fermions), the leading term in Eq.(\ref{tunnel_noneq}) is given by the equilibrium contribution. To account for the non-equilibrium situation in the electron case, one thus needs to compute Eq.(\ref{tunnel_noneq}) at the next leading order, which is proportional to the source emission probabilities $T_S$. Considering the case where one keeps identical emission probabilities $T_1=T_2=T_S$, but applies different voltage biases $V_1 \neq V_2$ at the input of QPC1 and QPC2 in order to tune the imbalance $I_-/I_+$, one has:
\begin{eqnarray}
P_{e}(I_-/I_+) & = & -T_S \frac{I_-}{I_+}, \label{Pe}
\end{eqnarray}
where $I_-/I_+$ is simply $(V_1-V_2)/(V_1+V_2)$.
Eq.(\ref{Pe}) is a clear hallmark of fermion (electron) behavior in a collision, as opposed to the anyon case. $P_e$ goes to zero (even in the case $I_- \neq I_+$) when the emission probability is decreased down to the Poissonian limit ($T_S \ll 1$). In contrast, $P$ is independent of $T_S$ in the Poissonian regime in the anyon case. This can be interpreted as braiding effects being present even in the case where a single source is switched on, whereas fermion antibunching can only be present when the two sources are switched on, leading to an additional dependence on $T_S$ in the electron case.

\section{Edge structure and fractional charges at $\nu=1/3$ and $\nu=2/5$}

\begin{figure*}
\includegraphics[width=2\columnwidth,keepaspectratio]{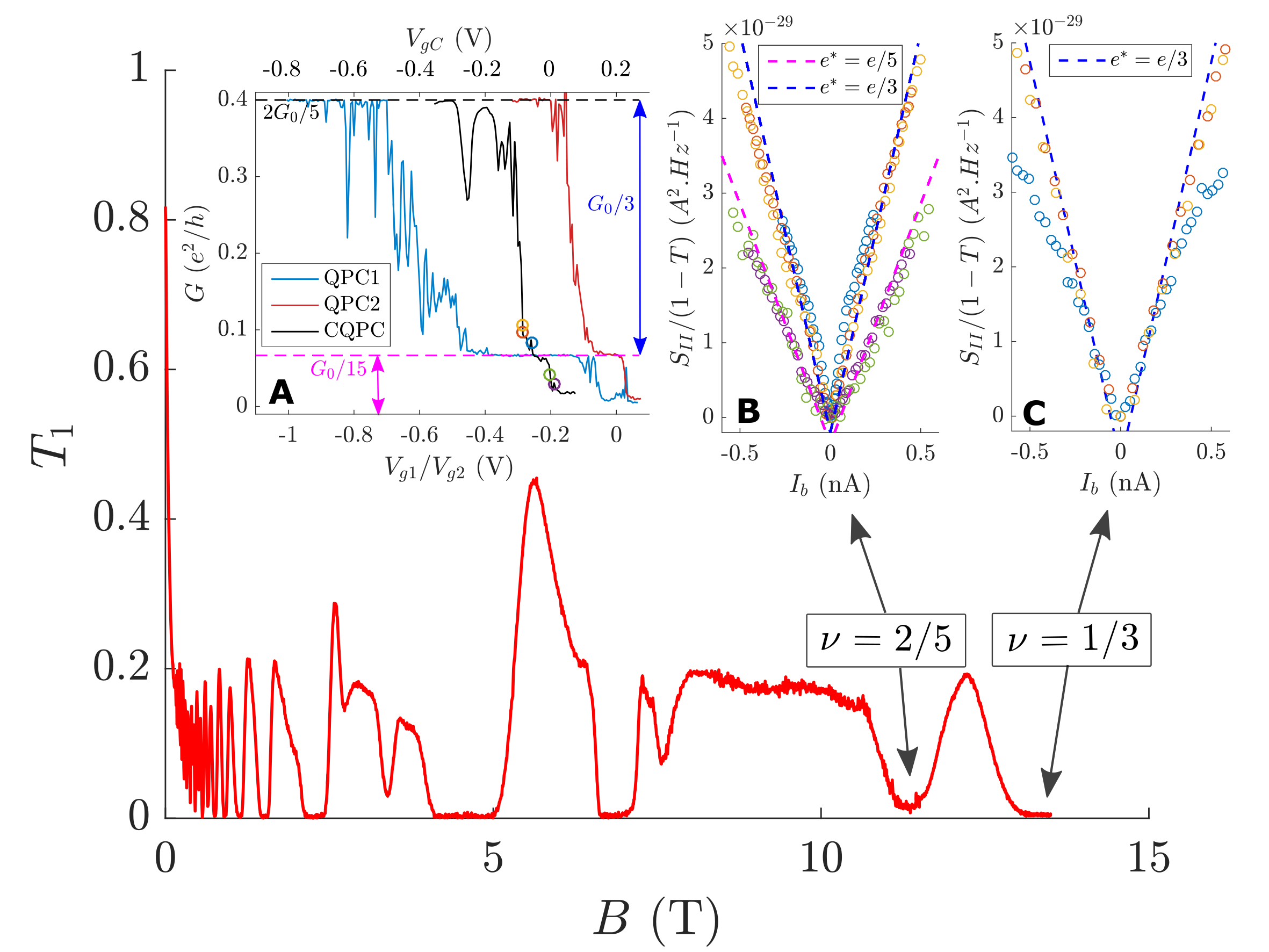}
\caption{\label{fig2} \textbf{Edge structure and fractional charges.} \\ Backscattering probability $T_1=\delta I_5/\delta I_7$ as a function of the magnetic field $B$. Inset \textbf{A}: Backscattering conductance $G(V=0)$ of QPC1, QPC2 and cQPC as a function of the QPC gate voltages at filling factor $\nu=2/5$. The colored circles correspond to the value of the conductance for the noise measurements plotted on insets \textbf{B}. Inset \textbf{B}:  Measurement of the noise $S_{II}$ at the output of cQPC normalized by the factor $(1-T)$ as a function of the backscattered current $I_b$ at filling factor $\nu=2/5$. Green and violet circles correspond to the partitioning of the inner channel whereas the blue, red and yellow circles correspond to the partitioning of the outer channel. The magenta dashed line is $ S_{II}/(1-T) =2e^*I_b$, with $e^*=e/5$ whereas the blue dashed line correspond to the fractional charge $e^*=e/3$. Inset \textbf{C}:  Measurement of the noise $S_{II}$ at the output of cQPC normalized by the factor $(1-T)$ as a function of the backscattered current $I_b$ at filling factor $\nu=1/3$. The blue circles correspond to $T(V=0)=0.14$, the red circles to $T(V)=0.29$, and the yellow circles to $T(V)=0.36$. The blue dashed line is $ S_{II}/(1-T) =2e^*I_b$, with $e^*=e/3$.}
\end{figure*}

The FQH phases can be identified by measuring the backscattering probability $T_1$ as a function of the magnetic field, see Fig.\ref{fig2}. As can be seen on the figure, the backscattering probability is suppressed each time the bulk becomes insulating and transport occurs at the edge. Backscattering is completely suppressed for $\nu=1/3$, whereas a small residual backscattering can be observed for $\nu=2/5$, with $T_1 \approx 0.03$. It is related to a slight depletion of the charge density at each QPC and it is not observed in bulk samples (without QPCs).

The edge structure at $\nu=2/5$ can be characterized by measuring the differential backscattering conductance of each QPC. It defined as the ratio of the small backscattered current $\delta I_b$ with the small a.c. voltage bias $\delta V$: $G= \delta I_b/\delta V$. Due to the non-linear I-V characteristics of QPCs in the fractional quantum Hall regime, the backscattering conductance $G(V)$ depends on the applied d.c. voltage bias $V$. The total backscattered current $I_b$ when a d.c. bias $V$ is applied can then be extracted from the measurement of $G(V)$ by $I_b=\int_0^{V} G(V') dV'$. $G(V=0)$ is plotted for the three QPCs in the inset A of Fig.\ref{fig2}. As discussed above, a small residual backscattering is present for positive gate voltages. Applying negative gate voltages, one observes a first conductance step at $G_{\text{in}}=G_0/15$ that corresponds to the backscattering of the inner channel. Applying a more negative voltage, one observes a second conductance step of amplitude  $G_{\text{out}}=G_0/3 $ that corresponds to the backscattering of the outer channel. Collision experiments will be performed either by setting all QPCs to backscatter the inner channel (see Fig.\ref{fig1}) or by setting all QPCs to backscatter the outer one. In this two-channel configuration, the backscattering probability is thus defined for each channel, with  $T_{\text{in}}= G/G_{\text{in}}$ for the inner channel and $T_{\text{out}}=(G-G_{\text{in}})/G_{\text{out}}$ for the outer one. When a d.c. voltage $V$ is applied, the backscattered currents on the inner/outer channels are thus given by: $I_{b,\text{in/out}}=G_{\text{in/out}} \int_0^{V} T_{\text{in/out}}(V') dV'$.

As already mentioned, the two edge channels at filling factor $\nu=2/5$ carry two different species of anyons with different fractional charges. The anyon fractional charge can be measured from the proportionality of the current noise $S_{II}$ at the output of a QPC with the backscattered current $I_b$ (see Ref.\cite{Kane94} for the theoretical prediction and Refs.\cite{Saminadayar97,dePicciotto97} for the first experimental measurements of fractional charges from noise measurements). In order to measure the fractional charge of anyons tunneling at cQPC, we plot in Fig.\ref{fig2}.B the current noise $S_{II}$ \cite{Note2} at the output of cQPC as a function of $I_b$. In this single QPC configuration for noise measurement, it is important to suppress any backscattering at input QPC1 and QPC2. As some residual backscattering is present when QPC1 and QPC2 are open, the experiment is performed by closing completely QPC1 and QPC2, and by sending a noiseless current $I_1^0$ towards cQPC (by applying a d.c. voltage $V_1$ at ohmic contact 1).

We first measure the anyon fractional charge on the inner channel, by setting cQPC to backscatter the inner channel (violet and green circles on  Fig.\ref{fig2}.B), and by measuring the evolution of $S_{II}$ with the backscattered current on the inner channel $I_{b,\text{in}}$. In order to take into account deviations from the Poissonian limit $T_{\text{in}} \ll 1$, we divide the noise $S_{II}$ by the usual factor $(1-T_{\text{in}})$ (see Refs.\cite{Trauzettel04,Reznikov99,Griffiths00,Feldman17}). The measurements are performed for two different values of $T_{\text{in}}(V_1=0)$, which corresponding conductance values $G(V_1=0)=G_{\text{in}}T_{\text{in}}(V_1=0)$ are represented by the violet and green circles in Fig.\ref{fig2}.A. The measurements of $S_{II}/(1-T_{\text{in}})$ are plotted in Fig.\ref{fig2}.B. As can be seen on the figure, the violet and green circles agree very nicely with the magenta dashed line which represents $S_{II}/(1-T_{\text{in}})=2e^* I_b$, with $e^*=e/5$. We next measure the anyon fractional charge on the outer channel for three different values of the backscattering probability of the outer channel $T_{\text{in}}(V_1=0)$ (the three corresponding values of $G(V_1=0)$ are represented by the blue, red and yellow circles on Fig.\ref{fig2}.A). The measurements of $S_{II}/(1-T_{\text{out}})$ as a function of $I_{b,\text{out}}$ are also plotted in Fig.\ref{fig2}.B (blue, red and yellow circles). They agree very well with the blue dashed line which represents $S_{II}/(1-T_{\text{out}})=2e^* I_b$, with $e^*=e/3$. Later on, two different configurations for anyon collisions will be studied at $\nu=2/5$, offering the possibility to probe the fractional statistics of two different species of anyons and to compare them. By setting QPC1, QPC2 and cQPC to partition the inner channel, we will realize the collision between anyons of charge $e/5$. When setting the all QPCs to partition the outer channel, we will realize the collision of anyons of charge $e/3$. As from now on it will be explicit which edge channel is backscattered by all QPCs (the outer or the inner), we will drop the indices $\text{out}$ and $\text{in}$ labeling the different transmissions in the rest of the manuscript in order to simplify the notations.

The same characterization can be performed for the filling factor $\nu=1/3 $. The backscattered conductance (not plotted here) resembles the conductance step of the outer channel at $\nu=2/5$, with a single conductance step of $G_0/3$ (as expected for a single edge channel). The noise measurements $S_{II}/(1-T)$ as a function of the backscattered current are plotted in Fig.\ref{fig2}.C for three different values of the backscattering transmission $T(V_1=0)=0.14$ (blue circles), $T(V_1=0)=0.29$ (red circles), and $T(V_1=0)=0.36$ (yellow circles). The measurements show strong similarities with those performed on the outer channel at $\nu=2/5$ with an agreement with the blue dashed line representing the scattering of anyons of fractional charge  $e/3$. However, contrary to the $\nu=2/5$ case, some deviations are observed at the largest values of the current for the smallest value of $T$ (blue circles).

The preliminary experiments described above confirm the edge structure at filling factor $\nu=1/3$ and $\nu=2/5$, with the expected species of anyons tunneling at cQPC in each case. We  now move to the description of anyon collisions, starting with the Laughlin $\nu=1/3$ case.

\section{The balanced collider in the Laughlin state $\nu=1/3$}

We start by presenting the measurements of the anyon collider in the balanced configuration ($I_-=0$) at the filling factor $\nu=1/3$. Here the edge structure is simple, and the topological order is characterized by a single number, which determines both the anyon fractional charge and the fractional statistics. It is therefore suitable as a reference state for extensive comparisons with the quantum model of anyon collisions, first developed in Ref.\cite{Rosenow16} followed by Refs.\cite{Morel22,Lee22}. We will focus firstly on the measurements of $P$ and on their comparison with predictions, taking $\lambda=\delta=1/3$ as a natural guess for evaluating Eq.(\ref{P}). These measurements have some similarities with the ones presented in Ref.\cite{Bartolomei20} but are extended here to a wider range of the values for cQPC's backscattering transmission $T$. They will also be used to compare, on the same sample, the different species of anyons at filling factor $\nu=2/5$ and $\nu=1/3$. Additionally, we also discuss our measurements of $T$ as a function of the anyon current $I_+$, and compare them with the quantum model. This is of particular interest as the characteristic non-linear evolution of $T$ with $I_+$ is predicted to follow a power law at high current $I_+$ with an exponent $2\delta -2$ \cite{Rosenow02}. It is reminiscent of the non-linear I-V characteristics in the tunneling current of a chiral Luttinger liquid \cite{Wen91,Kane96}. These measurements thus provide an independent probe of the value of $\delta$ in the anyon collisions.

\subsection{Measurements of the Fano factor $P$}

In order to measure $P(I_-=0)$ in the anyon collisions, we set $T_1$ and $T_2$ to almost identical values $T_1=T_2=T_S\approx0.05$. These small values of  emission probabilities are well within the Poissonian limit of anyon emission in the weak backscattering regime (anyon collisions for larger values of $T_S$  are discussed in Appendix A). We then measure the output current cross-correlations $S_{34}$ for different values of $T$ as a function of the total anyon current $I_+$ incoming on cQPC. Importantly, the tunneling charge at cQPC (plotted in Fig.\ref{fig2}.C) is constant (and equals $e^*=e/3$) on this large range of values of $T$.

\begin{figure*}
\includegraphics[width=1.6
\columnwidth,keepaspectratio]{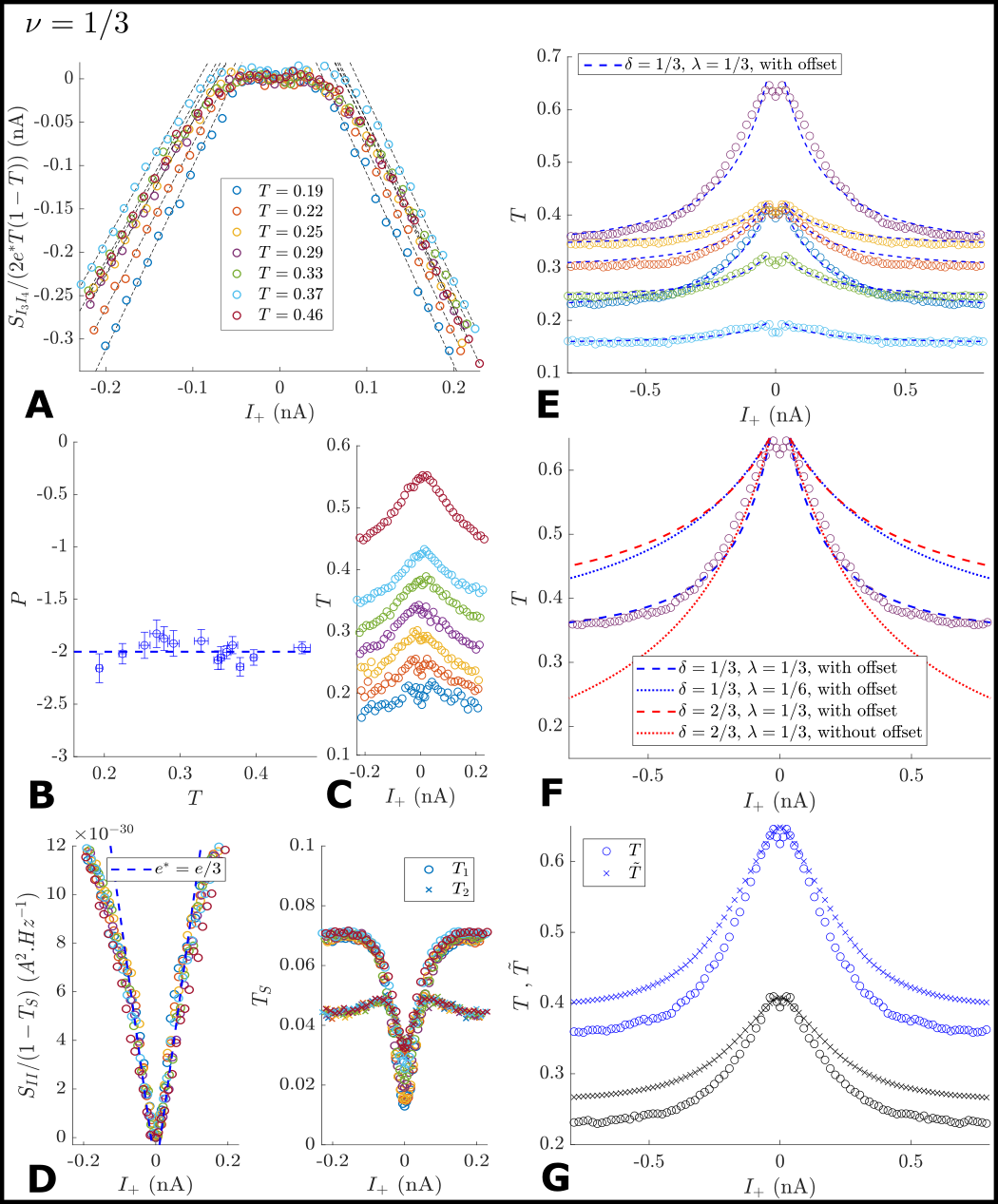}
\caption{\label{fig3}
\textbf{Balanced collider at $\nu=1/3$.} \\ \textbf{A} $S_{34}/\big[2e^*T(1-T)\big]$, with $e^*=e/3$ as a function of $I_+$ for different values of $T$ (plotted on panel \textbf{C}). The dashed lines are linear fits which slope $P$ is plotted on panel \textbf{B}. \textbf{B} $P$ (extracted from the linear fits of panel \textbf{A}) as a function of $T$. \textbf{C} Measurements of $T$ as a function of $I_+$ in the collision experiment.  \textbf{D} Left panel, $S_{II}/(1-T_S)$ as a function of $I_+$, with $S_{II}=S_{33}+ S_{44}+2S_{34}$. The blue dashed line is $2e/3 I_+$. Right panel, $T_1$ and $T_2$ as function of $I_+$. \textbf{E} $T$ as a function of $I_+$ in a larger range of input anyon currents and for different values of cQPC gate voltage. The blue dashed lines are comparisons with Eq.(\ref{T}), adding an offset of $35$~pA in $I_+$ that takes into account the thermal rounding at low $I_+$, and an offset $T_0$ in the values of $T$. The values of $T_0$ are extracted from a fit of the experimental data (see manuscript text) with $T_0 =0.21$ (blue points),  $T_0 =0.30$ (red points),  $T_0 =0.34$ (yellow points), $T_0 =0.33$ (violet points),  $T_0 =0.24$ (green points), and  $T_0 =0.16$ (cyan points). The other parameters for comparing data with model are $\lambda=\delta=e^*/e=1/3$ and $T_{\text{el}}=30$~mK.  \textbf{F} Comparison between data and Eq.(\ref{T}) using $\delta=\lambda=1/3$ and $T_0=0.33$ (blue dashed line), $\delta=1/3$, $\lambda=1/6$ and $T_0=0.33$ (blue dotted line), and $\delta=2/3$, $\lambda=1/3$ and $T_0=0.33$ (red dashed line), and using $\delta=2/3$, $\lambda=1/3$ and $T_0=0$ (red dotted line). \textbf{G}  Comparisons between the evolution of $T$ (circles) and $\tilde{T}$ (crosses) with $I_+$ at filling factor $\nu=1/3$ for two different values of cQPC gate voltage. }
\end{figure*}

The measurements of $S_{34}$ normalized by $2e^*T(1-T)$ are plotted in Fig.\ref{fig3}.A. All the measurements for different values of $T$ exhibit the same behavior with $S_{34} \approx 0$ at low current $|I_+|<40$~pA, followed by a linear variation with negative slope $P$ for larger values of $|I_+|$. Fig.\ref{fig3}.B presents the measurements of the slope $P$, extracted from linear fits of the data plotted in Fig.\ref{fig3}.A, as a function of $T$. The values of $P$ are remarkably constant in the large range $0.15 \leq T \leq 0.45$, meaning that the $T$ dependence of $S_{34}$ is perfectly captured by the factor $T(1-T)$. It shows that the determination of $P$ can be extended beyond the weak-backscattering regime for cQPC by taking into account deviations from $T\ll 1$ by the usual $(1-T)$ factor. The measured values of $P$ also agree very well with the predicted value $P=-2$ for anyons of exchange phase $\varphi=\pi/3$ at filling factor $\nu=1/3$, using $\lambda=\delta=1/3$ in Eq.(\ref{P}).

Fig.\ref{fig3}.C represents the evolution of $T$ as a function of $I_+$ for all the collision data plotted on panel A. All the data show the same qualitative behavior, with $T$ decreasing when $I_+$ increases. This is the analog, in a collision experiment, of the characteristic non-linear evolution of the backscattered current with applied voltage in single QPC tunneling experiments (see Refs.\cite{Roddaro03,Roddaro04} for $\nu=1/3$ and Refs.\cite{Radu08, Lin12, Baer14, Fu16} for $\nu=5/2$). A detailed analysis of this non-linear evolution is performed in a larger range of $I_+$ in the following section.

Finally Fig.\ref{fig3}.D (left panel) represents the simultaneous determination of the fractional charge $e^*$ emitted by QPC1 and QPC2. It is extracted from the measurement of the total noise $S_{II}=S_{33}+ S_{44}+2S_{34}$. From the conservation of the current between the inputs and outputs of cQPC, one has $S_{II}=S_{11}+S_{22}$, which is the noise generated from the random emission of anyons at QPC1 and QPC2. It is related to the anyon charge $e^*$ via the usual relation $S_{II}=2e^*(1-T_S) I_+$, where $I_+$ is the anyon current backscattered at QPC1 and QPC2 and where the factor $(1-T_S)$ takes into account deviations from the Poissonian limit (here, as $T_S \approx 0.05$, $1-T_S\approx 1$). As can be seen on Fig.\ref{fig3}.D, the data agrees well with the anyon charge $e^*=e/3$ (with some deviations around the maximum current of $200$~pA), showing that the charge measurements are consistent with the collision data. The right panel of Fig.\ref{fig3}.D represents the measurement of the anyon emission probabilities $T_1$ and $T_2$ as a function of $I_+$. There are slight differences between $T_1$ and $T_2$ leading to small deviations to the perfectly balanced case, with $I_-/I_+ =0.14 \pm 0.02$. However, such small variation of $I_-/I_+$ lead to very small variations of $P$ compared to its value $P(I_-=0)$ in the perfectly balanced case (the predicted difference between $P(I_-=0.14)$ and $P(I_-=0)$ is less than $2\%$,  smaller than error bars).

\subsection{Non-linear evolution of the backscattering transmission $T$}

Interestingly, the non-linearities of the I-V characteristics in tunneling experiments in the fractional quantum Hall regime are predicted to be related to the parameter $\delta$ \cite{Wen91,Fendley95,Kane96}. When considering anyon tunneling across a single QPC, the backscattering transmission  in the weak backscattering
regime is expected to decrease down to zero with a characteristic power law dependence, $T \propto V^{2\delta -2}$, where $V$ is the bias voltage across the QPC. However, previous experiments failed to provide a quantitative agreement with this prediction. In particular, experiments show a saturation of $T$ at large voltage to a constant value, or offset, that is different from 0. Subtracting this offset, many experiments have observed a quantitative agreement with predictions for the dependence of $T$ with the applied voltage. For example, in the $\nu=5/2$ case, by fitting the experimental data (after removal of the offset) with the $T(V)$ dependence  predicted by the model, several works \cite{Radu08, Lin12, Baer14, Fu16} have been able to extract the anyon fractional charge $e^*$ and the tunneling exponent $\delta$, and to use these values to discriminate between different possible topological orders of the $\nu=5/2$ state.

Quantum models of anyon collisions \cite{Rosenow16,Lee22} also provide quantitative predictions for the evolution of $T$ with $I_+$. They differ from the non-linear behavior predicted in single QPC experiments (in particular in collision experiments, the backscattered current at cQPC evolves non-linearly with the input anyon current $I_+$, instead of the voltage $V$ in single QPC experiments). However, they share common features, with $T$ predicted to decrease down to zero at large current $I_+$ with a power law of the same exponent $2\delta -2$. Interestingly, the exact dependence of $T$ with $I_+$ is richer than the power law, as it  depends on all the collision parameters $e^*$, $\lambda$ and $\delta$ (see Appendix A):

\begin{eqnarray}
 T & = & \alpha f(\delta,\xi_+) \label{T} \\
 f(\delta,\xi_+) &= & \frac{\Gamma(\delta + \xi_+)}{\Gamma(1-\delta + \xi_+) } \big[\Psi(1-\delta + \xi_+)-\Psi(\delta + \xi_+)\big] \\
  \xi_+ & = & \frac{\hbar I_+}{2 \pi e^* k_B T_{\text{el}}} \big[1-\cos{(2\pi \lambda)}\big]
\end{eqnarray}
where $\Gamma$ and $\Psi$ are the gamma and digamma functions (with $\Psi(x)=\Gamma'(x)/\Gamma(x)$), and $\alpha$ is a constant which does not depend on $I_+$.

Fig.\ref{fig3}.E shows the measurements of $T(I_+)$  in a large range of currents $I_+$, in order to grasp the complete non-linear dependence of $T$ with the anyon current. For better signal over noise ratio and to suppress a small asymmetry of $T$ between positive and negative values of $I_+$, we average together $T(I_+)$ and $T(-I_+)$ and plot on Fig.\ref{fig3}.E the symmetrized transmission $(T(I_+)+T(-I_+))/2$. As shown on the figure, the data shows the expected decrease of $T$ with $I_+$. However, as observed in single QPC experiments, $T$ does not decrease down to zero but saturates at large currents at an offset value. Following the analysis performed in single QPC experiments \cite{Roddaro03,Roddaro04,Radu08, Lin12, Baer14, Fu16}, we compare our experimental data with the predictions of Eq.(\ref{T}) up to an offset value $T_0$ that is added to Eq.(\ref{T}) in order to match the experimental data. The offset is extracted from a fit of the experimental data by a power law, $T(I_+)=T_0+I_+^\alpha$, where $T_0$ and $\alpha$ are the two fit parameters. As the power law is the asymptotic limit of Eq.\ref{T} for large currents, the fit is restricted to $|I_+| \geq 200$~pA. The typical uncertainty on the extracted values of the offset is $\pm 5\times 10^-3$.  Eq.(\ref{T}) is also only valid for currents larger than the thermal limit ($I_+/e^* \geq k_B T_{\text{el}}/h$, see Appendix A), therefore it cannot capture the thermal rounding observed on the experimental curves for $|I_+| \leq 35$~pA. In order to take into account these thermal effects, we add an offset of $35$~pA to the values of $I_+$ in Eq.(\ref{T}) (the same offset on $I_+$ is used on all the traces). Finally, we fix the values of all the other parameters in Eq.(\ref{T}), with $\lambda=\delta=1/3$, $T_{\text{el}}=30$~mK and $\alpha=T(I_+=0)/f(\delta,\xi_+=0)$. The blue dashed lines on the figure represent the result of this analysis. As can be seen on Fig.\ref{fig3}.E, the agreement with the data is excellent.

Fig.\ref{fig3}.F shows additional comparisons with Eq.(\ref{T}). The red dashed line and the blue dotted line represent the same analysis modifying the value of one parameter. For the red dashed line, the value of $\delta$  is modified to $\delta=2/3$, whereas for the blue dotted line, the value of $\lambda$ is modified to $\lambda=1/6$. For these choices of values of $\delta$ and $\lambda$, the agreement is very poor, showing that the analysis discriminates between different values of the couples of values $(\delta, \lambda)$. The red dotted line corresponds to the same analysis without adding the offset value to  Eq.(\ref{T}), and choosing $\delta=2/3$ and $\lambda=1/3$. It illustrates the crucial role played by the offset for comparisons between data and model, as a good agreement is observed with this choice of parameters only for low values of the current $I_+ \leq 250$~pA. By contrast, the agreement becomes very poor beyond this value, as this analysis does not capture the saturation of $T$ at a finite value.

The collision data discussed above provide a consistent agreement with the predictions for anyons at filling factor $\nu=1/3$. In particular, analyzing together the evolution of current cross-correlations $S_{34}$ and the backscattering transmission $T$ with the anyon current $I_+$ supports $\lambda=\delta=1/3$ for the choice of parameters. However, two discrepancies with the model remain to be explained. The first one, discussed above, is the saturation of the backscattering transmission which does not decrease down to zero at high anyon current $I_+$. The second one is related to the two different experimental schemes that can be used to measure the backscattering transmission $T$, depending on which ohmic contact is used to inject the electrical current in the sample. In the first scheme, discussed so far, the small a.c. current is injected via ohmic contact $2$, leading to the following expression for $T$: $T=\frac{\delta I_3}{T_2 \delta I_2^0}$. This is the definition introduced in Ref.\cite{Rosenow16}. In the second scheme, the small a.c. current $\delta I_8$ is injected on contact $8$, moving the a.c. source $V_{\text{ac}}$ (see Fig.\ref{fig1}) from contact $2$ to contact $8$. Labeling $\tilde{T}$ cQPC's backscattering transmission measured in this experimental scheme, one has $\tilde{T}=\frac{\delta I_3}{\delta I_2}=\frac{\delta I_3}{(1-T_2) \delta I_8}$. Due to the non-linearities of each QPC, $T$ and $\tilde{T}$ are predicted to be different in the limit of large currents $I_+$ (with $T$ smaller than $\tilde{T}$), with a ratio governed by the parameter $\lambda$ \cite{Lee22}. As shown in Fig.\ref{fig3}.G, we have a qualitative agreement with this prediction, with $T$ and $\tilde{T}$ reaching different values at large $I_+$, with $T$ smaller than $\tilde{T}$. However, the quantitative agreement is poor, as the difference reaches only $15\%$, much smaller than $\approx 50\%$ predicted for $\lambda=1/3$. As a consequence, the measurement of $P$, when the current range is restricted to $|I_+| \leq 300$~pA, depends only weakly on the definition of $T$ that is used for the normalization of the cross-correlations: $S_{34}/\big[T(1-T)\big] \approx S_{34}/\big[\tilde{T}(1-\tilde{T})\big]$ within variations smaller than $10\%$.

\section{The balanced collider at the filling factor $\nu=2/5$} \label{Bal_2_5}

We now turn to the experimental study of anyon collisions at the filling factor $\nu=2/5$. Due to the more complex structure of the edge channels, with two co-propagating channels at $\nu=2/5$, we can probe two different species of anyons depending on the chosen experimental configuration. We first consider the partitioning of the outer channel by all QPC, thereby implementing a collision between anyons of fractional charge $e^*=e/3$ generated by QPC1 and QPC2. As shown below, the collision results in this case have strong similarities with the one performed on the Laughlin state $\nu=1/3$. We then consider the partitioning of the inner channel by all QPCs. This situation is the most interesting one, as we probe a different variety of anyons, with a different fractional charge $e^*=e/5$ and carried by a channel of conductance $G_{\text{in}}=G_0/15$. As shown below, we observe clear quantitative differences with the $\nu=1/3$ case, with a reduced value of $|P|$ which can be interpreted as a reduced level of anyon bunching.

\subsection{Collisions of anyons of fractional charge $e^*=e/3$}

\begin{figure*}
\includegraphics[width=1.6
\columnwidth,keepaspectratio]{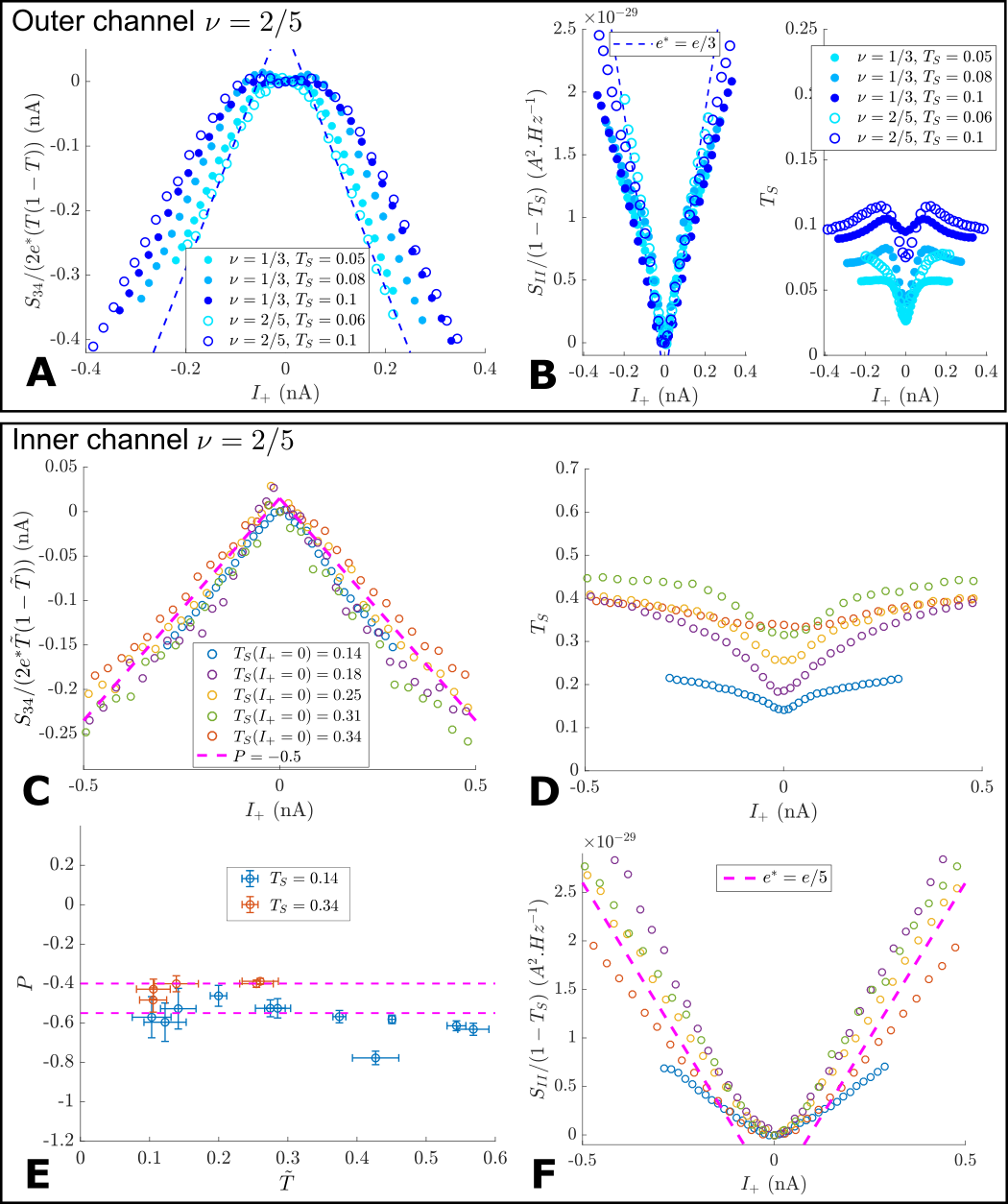}
\caption{\label{fig5} \textbf{Balanced collider at $\nu=2/5$, outer and inner channels.} \\ \textbf{A} $S_{34}/\big[2e^*T(1-T)\big]$, with $e^*=e/3$ as a function of $I_+$ for different values of the emission probability $T_S$.  Measurements on the outer channel of $\nu=2/5$ (blue circles) are compared with the measurements at $\nu=1/3$ (filled blue circles).  \textbf{B}  Left panel $S_{II}/(1-T_S)$ as a function of $I_+$, with $S_{II}=S_{33}+ S_{44}+2S_{34}$. The blue dashed line is $2e/3 I_+$. Right panel, $T_1$ and $T_2$ as function of $I_+$.  \textbf{C}  $S_{34}/\big[2e^*\tilde{T}(1-\tilde{T})\big]$, with $e^*=e/5$ as a function of $I_+$. Measurements are performed on the inner channel of $\nu=2/5$ for different values of the emission probability $T_S$. Data points averaged on several values of cQPC's backscattering transmission $\tilde{T}$.  \textbf{D} Measurements of $T_S$ as a function of $I_+$. \textbf{E} $P$ (extracted from linear fits of the normalized cross-correlations) as a function of $\tilde{T}$ for two different values $T_S$. \textbf{F} $S_{II}/(1-T_S)$ as a function of $I_+$, with $S_{II}=S_{33}+ S_{44}+2S_{34}$. The magenta dashed line is $2e/5 I_+$.  }
\end{figure*}

In order to implement a collision between anyons of fractional charge $e^*=e/3$, we set all QPCs (QPC1, QPC2 and cQPC) to partition the outer edge channel. For a better separation between the partitioning of the inner and outer channels for all QPCs, we also set the magnetic field slightly away from the minimal value of the backscattering probabilities at $\nu=2/5$ (see Fig.\ref{fig2}) so as to perfectly reflect the inner channel of $\nu=2/5$ at each QPC ($T_{\text{in}}=1$). Fig.\ref{fig5}.A presents the measurements of the current cross-correlations $S_{34}$ normalized by the factor $2e^*T(1-T)$, with $e^*=e/3$, as a function of the anyon current $I_+$. Data points for the collision performed on the outer channel of $\nu=2/5$ are represented by blue circles (for two transmissions $T_S=0.06$ and $T_S=0.1$). They are compared with data points at filling factor $\nu=1/3$, which are represented by filled blue circles (for $T_S=0.05$, $T_S=0.08$, and $T_S=0.1$). One can see that all traces look very similar, with small variations of the measured slope $P$ (the $T_S$ values and their variation with $I_+$ are shown on Fig.\ref{fig5}.B, right panel). For $\nu=1/3$, one has $P=-1.92 \pm 0.15$ for $T_S=0.05$,  $P=-1.87 \pm 0.15$ for $T_S=0.08$ and $P=-1.67 \pm 0.1$ for $T_S=0.1$. These results support the fact that $P$ is almost independent of $T_S$ in the regime of Poissonian anyon emission, with $P$ varying by $\approx 13\%$ when $T_S$ varies by a factor 2. However, increasing $T_S$ away from the Poissonian regime leads to a decrease of $|P|$ from its maximum values reached at small values of $T_S$. The results are very similar at filling factor $\nu=2/5$, with $P=-1.95 \pm 0.15$ for $T_S=0.06$ and $P=-1.61 \pm 0.09$ for $T_S=0.1$, confirming the decrease of $|P|$ when $T_S$ increases.  Consistently with the similar behavior observed in anyon collisions, one can see that the measurements of the charge emitted by QPC1 and QPC2 (plotted in Fig.\ref{fig5}.B, left panel) are also similar. All the measurements of $S_{II}=S_{33}+S_{44}+ 2S_{34}$, normalized by the factor $(1-T_S)$ (always $\leq 10\%$ for the values of $T_S$ discussed here), agree with the expected value for charge $e^*=e/3$ (blue dashed line). The agreement is even better at large currents $I_+$ for the outer channel of $\nu=2/5$ than for $\nu=1/3$.

These measurements are consistent with the usual picture describing the outer edge channel of $\nu=2/5$ as an effective $\nu=1/3$ fractional quantum Hall state, with the same conductance $G_0/3$ and carrying the same anyons of charge $e^*=e/3$. They also provide an additional demonstration of the robustness of anyon collision signals in the simple case $\varphi/\pi=e^*/e=1/3$.

\subsection{Collisions of anyons of fractional charge $e^*=e/5$}

We now move to the collision experiments between anyons of charge $e^*=e/5$, by setting all the QPCs to backscatter the inner channel of $\nu=2/5$. The conductance of the inner channel being five times smaller than the conductance of the outer one, our measurements of $T$ in the regime of weak-backscattering of all QPCs are not accurate enough to capture well the evolution of $T$ with $I_+$. As discussed for $\nu=1/3$, the differences between $T$ and $\tilde{T}$ are small at low currents, $|I_+| \leq 400$~pA, leading to small differences ($\leq 10\%$ ) between the slopes of $S_{34}/\big[2e^*T(1-T)\big]$ and  $S_{34}/\big[2e^*\tilde{T}(1-\tilde{T})\big]$. As our measurements of $\tilde{T}$ are more accurate than the measurements of $T$ for the inner channel of $\nu=2/5$, we analyze here the measurements of $S_{34}/\big[2e^*\tilde{T}(1-\tilde{T})\big]$ for better accuracy.

Fig.\ref{fig5}.C presents the measurements of the current cross-correlations normalized by the factor $2e^*\tilde{T}(1-\tilde{T})$, using $e^*=e/5$, as a function of $I_+$. The measurement data for five different values of the anyon emission probability $T_S$ at $I_+=0$ are plotted, with $T_S$ ranging from $T_S \approx 0.15$ to $T_S \approx 0.45$ (the variations of $T_S$ with $I_+$ are plotted in Fig.\ref{fig5}.D). As can be seen in Fig.\ref{fig5}.C, all data show the same behavior with a linear evolution of the normalized cross-correlations with a negative slope that is characteristic of a collision involving anyons. If the qualitative behavior of anyons of charge $e^*=e/5$ is similar to that of anyons of charge $e/3$, there is a strong quantitative difference. $P$ is much smaller for $\nu=2/5$, with $P \approx -0.5$ (represented by the magenta dashed line in Fig.\ref{fig5}.C). As can also be seen on the figure, all data for different emission probabilities $T_S$ show small variations of the slope. As in the $\nu=1/3$ case, the largest value of $|P|$ is obtained for the smallest value of $T_S$ with $P=-0.57 \pm 0.02$ for $T_S=0.14$ (blue circles). Deviations of the measured values of $P$ for different values of $T_S$ are rather small, with a minimum value of $P=-0.43 \pm 0.02$ for $T_S=0.34$ (red circles). The mean value of $P$ averaged on all values of $T_S$ is $P=-0.51$ with a standard deviation of $0.06$.

Fig.\ref{fig5}.E shows the measurement of the slope $P$ as a function of cQPC backscattering transmission $\tilde{T}$ for two different emission probabilities $T_S=0.14$ (blue circles) and $T_S=0.34$ (red circles), which correspond to the maximum and minimum measured values of $P$. As in the $\nu=1/3$ case, the normalization factor $\tilde{T}(1-\tilde{T})$ captures well the variation of the current cross-correlations with $\tilde{T}$. The measured values of $P$ show very small deviations when varying $\tilde{T}$. Exploiting this small variation of the measured slopes with $\tilde{T}$ for better signal to noise ratio, we average together in Fig.\ref{fig5}.C several measurements carried out for different values of $\tilde{T}$.

For consistency, we plot in Fig.\ref{fig5}.F the measurement of the fractional charge generated by QPC1 and QPC2 deduced from the measurement of $S_{II}$. As expected for the partitioning of the inner channel at $\nu=2/5$, the measurements are consistent with the charge $e^*=e/5$ (magenta line), except the measurement for the most open value of QPC1 and QPC2, $T_S=0.14$. This discrepancy may be related to the residual backscattering that is observed when all QPCs are open at their maximum value (corresponding to $T_S \approx 0.14$).

These results of collision experiments carried out on anyons of fractional charge $e^*=e/5$ show that anyon colliders provide clear quantitative differences between different species of anyons, with different fractional charge and different statistics. The difference between the measured values of $P$ for $\nu=1/3$ and $\nu=2/5$ is much larger than differences in the fractional charge, with a factor $4$ for the values of $P$, compared to a factor $1.66$ for the charge. Note that the ratio of the fractional charges is already taken into account in the definition of $P$, meaning that the ratio of the slopes of the negative cross-correlation $S_{34}$ is a factor $6.7$, largely exceeding the difference between the anyon's fractional charges.

Comparisons with the predictions of Eq.(\ref{P}) are more complex in the $\nu=2/5$ case than at $\nu=1/3$ as Ref.\cite{Rosenow16} does not address explicitly the case where several co-propagating channels are present. Nonetheless, one may compare our experimental results with Eq.(\ref{P}) using two naive guesses for the couple of values $(\delta,\lambda)$. Regarding first the value of $\delta$, the predicted exponent for the tunneling of quasiparticles of fractional charge $e/5$ at filling factor $\nu=2/5$ is $\delta=3/5$ \cite{Ferraro10}. Regarding the value of $\lambda$, two cases can be considered. In the first one,  $\lambda$ is related to the exchange phase of anyons in the bulk at filling factor $\nu=2/5$: $\lambda =3/5$ \cite{Lopez99}. Eq.(\ref{P}), with $\lambda=\delta=3/5$, predicts the surprising large positive value of $P(I_-=0)=+6$. This result has to be interpreted with caution: the cross-correlations $S_{34}$ are still predicted to be negative, but the backscattering transmission $T$ is also predicted to be negative, leading to positive values of $P$ (this is the case for all values of $\lambda >1/2$). Even though our measurements of  $T$  are less accurate than our measurements of $\tilde{T}$, negative values of $T$ can be ruled out, excluding the possibility to compare our experimental data with $\lambda=3/5$.

The second possible value of $\lambda$ is drawn of the quantum model of anyon collisions developed in Ref.\cite{Rosenow16}, where the charge density $\rho(x,t)$ in each edge channel at the input of cQPC is represented by a bosonic field $\phi(x,t)$, with $\rho(x,t)=\frac{e}{2 \pi} \frac{\partial \phi}{  \partial x}$. $2\pi \lambda$ is then related to the kick in the bosonic field $\Delta \phi$ when an anyon is emitted by an input QPC. This shows a direct relation between the charge carried by each anyon and $\lambda$ as $e^*= e\Delta \phi/(2\pi)=e \lambda$. A natural choice for $\lambda$ is thus related to the fractional charge, $\lambda=e^*/e=1/5$. Eq.(\ref{P}), with $\lambda=1/5$ and $\delta=3/5$, predicts $P(I_-=0)=-0.18$, which captures  the reduction of $|P|$ when moving from $\nu=1/3$ to $\nu=2/5$, even though the quantitative agreement is poor.

In order to provide more input for comparisons between data and models, we move in the next section to the study of the unbalanced collider, by varying the imbalance ratio $I_-/I_+$.

\section{The unbalanced collider, comparisons between $\nu=1/3$ and $\nu=2/5$} \label{Unbal_2_5}

The study of the anyon collider in the unbalanced case provides important additional manifestations of anyon braiding in the collision process. As such it provides both additional data to discriminate between anyons and fermions and between different species of anyons. As shown by Eq.(\ref{tunnel_noneq}), the first leading term responsible for the observation of negative cross-correlations in an anyon collision explicitly depends on the braiding factor via the difference $e^{\pm i 2\pi \lambda}-1$. As a consequence of the trivial character of the braiding phase ($e^{ i 2\pi \lambda}=1$) in the fermion case, this leading contribution vanishes for electrons, and non-equilibrium contributions have to be computed at the next order. As a result, $P$ is proportional to $T_1=T_2=T_S$ in the electron case, implying that $P$ goes to zero when $T_S$ goes to zero. This contrasts with the anyon case which shows the maximum values of $|P|$ for the smallest values of $T_S$ (see for example Fig.\ref{fig5}.A). Additionally, the braiding factor $e^{\pm i 2\pi \lambda}$ appears with opposite signs for anyons emitted by source 1 compared to those emitted by source 2 in Eq.(\ref{tunnel_noneq}). This means that the contribution of both sources is not additive and one expects interference contributions tuned by the ratio $I_-/I_+$. The exact dependence of $P$ on the imbalance ratio provides a stringent test to compare data with the models, and especially, to extract the couples of parameters $(\delta, \lambda)$ that agree best with collision signals for a given variety of anyons.

\begin{figure*}
\includegraphics[width=1.6
\columnwidth,keepaspectratio]{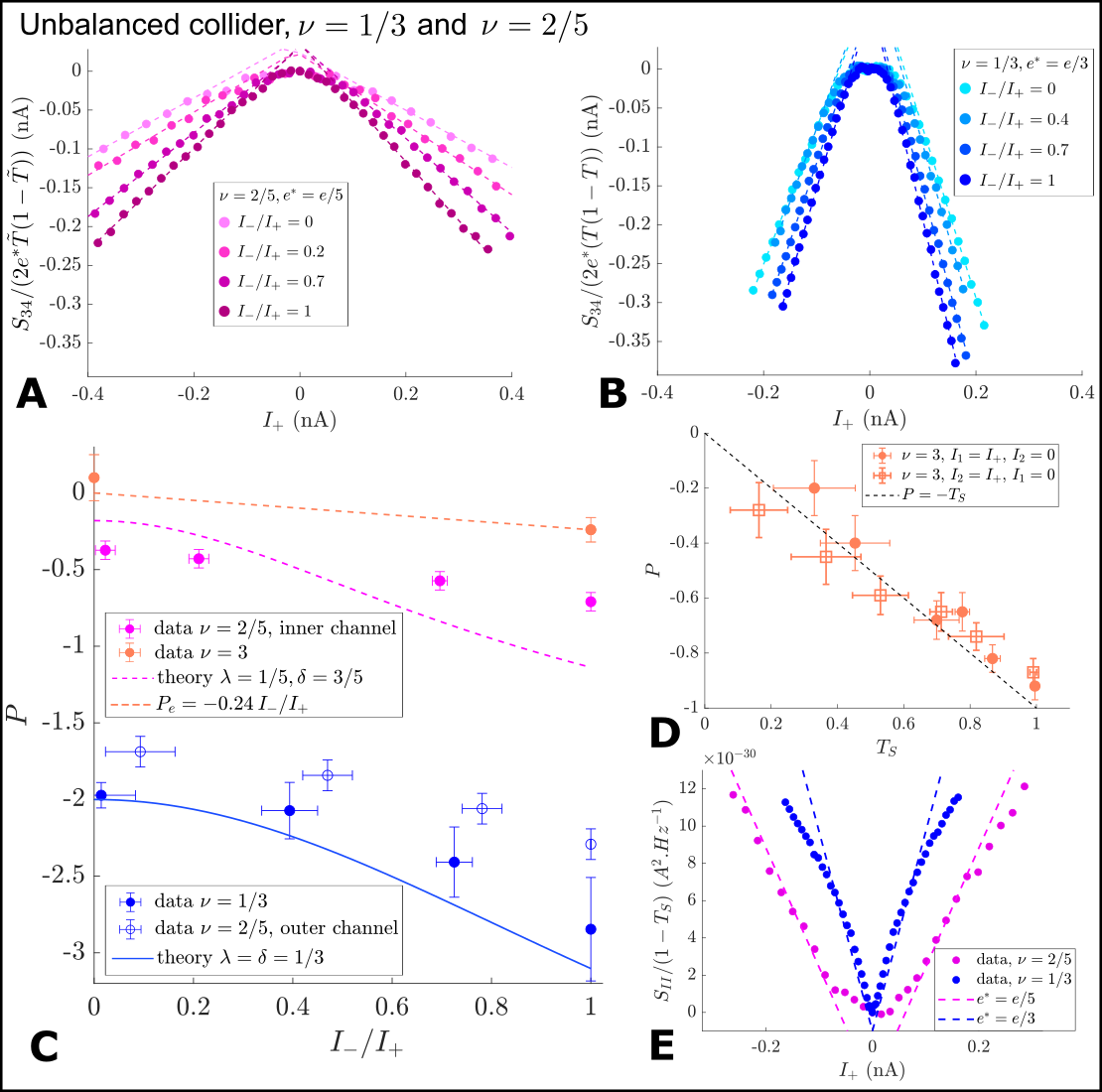}
\caption{\label{fig7}  \textbf{Unbalanced collider, comparison between $\nu=1/3$ and $\nu=2/5$.} \\ \textbf{A} $S_{34}/\big[2e^*\tilde{T}(1-\tilde{T})\big]$, with $e^*=e/5$ for the inner channel of $\nu=2/5$ as a function of $I_+$. The different plots correspond to different values of the imbalance ratio $I_-/I_+$.   \textbf{B}  $S_{34}/\big[2e^* T(1-T)\big]$, with $e^*=e/3$ for  $\nu=1/3$ as a function of $I_+$. The different plots correspond to different values of the imbalance ratio $I_-/I_+$. \textbf{C} $P$ as a function of $I_-/I_+$ for $\nu=1/3$ (filled blue circles), the outer channel of $\nu=2/5$ (blue circles), the inner channel of $\nu=2/5$ (filled magenta circles) and the electron case at $\nu=3$ (filled orange circles). The blue line is the prediction from Ref.\cite{Rosenow16} using $\lambda=\delta=1/3$. The magenta dashed line is the prediction using $\lambda=1/5$ and $\delta=3/5$, and the orange line is the prediction from Eq.(\ref{Pe}) with $T_S=0.24$.  \textbf{D} $P$ for $I_-=I_+$ as a function of $T_S$ in the integer case ($\nu=3$). The filled orange dots correspond to source 1 switched on (with source 2 switched off). The empty orange dots correspond to source 2 switched on (with source 1 switched off). The black dashed line is the prediction from Eq.(\ref{Pe}), $P=-T_S$. \textbf{E} Measurement of the fractional charge $e^*$ extracted from the ratio $S_{II}/(1-T_S)$ as a function of the anyon current $I_+$ (for $I_-=I_+$). The blue dashed line is the prediction for $e^*=e/3$ and the magenta dashed line the prediction for $e^*=e/5$.  }
\end{figure*}

We plot in Fig.\ref{fig7}.A the results of anyon collisions performed on the inner channel of $\nu=2/5$ for different values of the imbalance ratio $I_-/I_+$. The imbalance is tuned by changing the ratio $(V_1-V_2)/(V_1+V_2)$, while keeping the two emission probabilities equal, $T_1 = T_2 =T_S=0.35$. As one can see in Fig.\ref{fig7}.A, $|P|$ increases when the ratio $I_-/I_+$ is increased, in accordance with the expected interference between the two sources caused by anyon braiding.  The variation of the slope $P$ is large in relative value, with $P$ varying approximately by a factor $1.75$ when going from $I_-/I_+=0$ ($P \approx -0.4$) to $I_-=I_+$ ($P \approx -0.7$). Fig.\ref{fig7}.B presents the same measurements of the normalized cross-correlations for different imbalance ratios in the $\nu=1/3$ case. Here also we observe the characteristic increase of $|P|$ when $I_-/I_+$ increases, with  $P(I_-=0)\approx -2$ and $P(I_-=I_+) \approx -3$. These results for $\nu=1/3$ reproduce, in a different sample, those presented in Ref.\cite{Bartolomei20}.

The measurements of $P(I_-/I_+)$ are plotted in Fig.\ref{fig7}.C for filling factors $\nu=1/3$ and $\nu=2/5$ (outer and inner channels), as well as the integer case $\nu=3$ for comparison between anyons and electron. These measurements provide, on a single plot, the striking differences between electrons and anyons as well as quantitative way to distinguish between different species of anyons. Regarding first the $\nu=3$ case, one can clearly see that $P(I_-=0) \approx 0$ as a consequence of fermionic antibunching. ($P(I_-=0)$ is even slightly positive in the electron case.) Increasing the imbalance ratio towards $I_-=I_+$ restores the negative cross-correlations with $P(I_-=I_+) = -0.25 \pm 0.08$ for $T_S=0.24 \pm 0.05$. The suppression of fermionic antibunching for $I_-=I_+$ makes this situation a priori less convenient for the observation of clear difference between fermions and anyons, as cross-correlations are negative in both cases. However, the dependence of $P$ with $T_S$ for $I_-=I_+$ is completely different in the two cases. For electrons, the prediction $P(I_-=I_+)=-T_S$ shows a linear decrease of $|P|$ when $T_S$ decreases, with $P \approx 0$ in the Poissonian limit $T_S \ll 1$. This is exactly what we observe in Fig.\ref{fig7}.D at filling factor $\nu=3$ (see also Refs.\cite{Oliver99,Oberholzer00}). This contrasts completely with the increase of $|P|$ when decreasing $T_S$ observed in the anyon case (see Fig.\ref{fig5}.A).

Focusing now on the measurements of $P(I_-/I_+)$ in the anyon cases, all plots for $\nu=1/3$ and $\nu=2/5$ show an increase of $|P|$ when $I_-/I_+$ increases, as predicted. Due to the large scale of the figure, the decrease of $|P|$ is less apparent in the $\nu=2/5$ case (inner channel), even though it is larger in relative values. These plots can be compared with predictions of the anyon model. The $\nu=1/3$ case is the most straightforward, where a natural choice of $\lambda$ and $\delta$ is $\lambda=\delta=1/3$, which shows an excellent agreement with the data. Regarding the outer channel for $\nu=2/5$, the values of $P(I_-=0)$ are slightly smaller compared to $\nu=1/3$, which is probably related to the larger value of the anyon emission probability $T_S \approx 0.08$. The amplitude of variation of $P(I_-/I_+)$ with $I_-/I_+$ is also reduced by approximately $10\%$ compared with the $\nu=1/3$ case. This might be related to effects of interactions between the two co-propagating channels at $\nu=2/5$. These effects are more apparent in the case of the inner channel at $\nu=2/5$. This situation is clearly more open in terms of the choice of the parameters $\lambda$ and $\delta$ for comparisons with the models. As discussed before, there is no agreement between our data and the choice $\lambda=\delta=3/5$. The choice $\lambda=1/5$ and $\delta=3/5$ (magenta dashed line in Fig.\ref{fig7}.C) captures the right order of magnitude for $P(I_-/I_+)$, which is centered around $-0.6$. However, the amplitude of variation of $P$ with $I_-/I_+$ is much larger in the theoretical prediction compared to the experimental data. This shows that, contrary to the $\nu=1/3$ case, we do not obtain a quantitative agreement at $\nu=2/5$. The reduced contrast of the variations of $P$ with $I_-/I_+$ is reminiscent of the reduction of the contrast that is observed in electronic analogs \cite{Bocquillon13,Marguerite16} of the Hong-Ou-Mandel (HOM) experiment \cite{Hong87}. In these experiments, the interference between both sources is controlled by the time delay $\tau$ between the triggered emissions of single electrons by source 1 and source 2 of the collider. $\tau$ in electronic HOM experiments thus plays the role of $I_-/I_+$ in the anyon collider. In Ref.\cite{Marguerite16}, the reduction of the contrast was quantitatively captured by taking into account the Coulomb interaction between neighboring edge channels \cite{Wahl14,Ferraro14}. It suggests that the observed reduction of contrast at $\nu=2/5$ may also originate from interactions between the inner and the outer channel.

\section{Conclusion}

We have performed anyon collision experiments at two different filling factors $\nu=1/3$ and  $\nu=2/5$. Our results demonstrate the richness of anyon collisions, firstly to discriminate anyons from fermions, and secondly to distinguish between different species of anyons of different statistics.

Regarding the first point, anyons show common features at filling factor $\nu=1/3$ and $\nu=2/5$. As a result of anyon bunching, current cross-correlations are negative in the balanced case, with $P(I_-=0) <0$. In the unbalanced case, $|P|$ increases when $I_-/I_+$ increases, and importantly, $|P|$ increases when $T_S$ decreases, with non-zero values of $P$ in the Poissonian limit $T_S \ll 1$. This comes from the fact that braiding mechanisms are present even in the limit where a single source is switched on at the input of the collider. This contrasts completely with the fermion case, where one has  $P(I_-=0)=0$ as a result of fermion antibunching. In the unbalanced case, negative cross-correlations are restored, $P(I_-=I_+) <0$. But contrary to the anyon case, $|P|$ decreases when $T_S$ decreases, with $|P(I_-=I_+)|=T_S \ll 1$ in the Poissonian limit. This results from the fact that fermionic antibunching requires one electron to be emitted by each source, leading to an additional dependence in $T_S$ in the value of $P$.

Regarding the second point, the anyon collider shows robust experimental signatures for a given type of anyons and distinct signatures for different species of anyons. The experimental measurements of $P(I_-/I_+)$ reproduce similar values at $\nu=1/3$ on different samples and in independent experiments by different groups. Additionally, measurements performed on the outer channel of $\nu=2/5$ reproduce similar results, consistently with the predictions of anyon properties in this case. However, the dependence of $P$ with the imbalance ratio $I_-/I_+$ shows a slightly smaller contrast for the outer channel of $\nu=2/5$ than for $\nu=1/3$. This small deviation might be related to residual interchannel interaction effects in the $\nu=2/5$ case. Anyon collision experiments performed on the inner channel of $\nu=2/5$ show clear distinct experimental signatures, with a value of $P$ reduced by more than a factor $2$ compared to $\nu=1/3$. Our results point towards a reduced value of the parameter $\lambda \approx 1/5$ that would be close to the anyon fractional charge $e^*/e=1/5$ in this case. However, contrary to the $\nu=1/3$ case, the agreement with the model is not good. In particular, the amplitude of variation of $P$ with $I_-/I_+$ is strongly reduced compared to predictions. This suggests that other mechanisms, such as interactions between the two co-propagating edge channels, play an important role in this case. In this respect, it would be very interesting to compare these results with the $\nu=1/5$ case, which carries anyons of the same fractional charge, but with a simple edge structure. Studying anyon collisions for various propagation lengths between the input QPC and the central beam-splitter would also provide a quantitative probe of interaction effects \cite{LeSueur10}, as those are predicted to increase for increasing interaction distances. This rich phenomenology of anyon collisions for different abelian topological orders can also be extended to the non-abelian case \cite{Lee22}. In the former, braiding mechanisms show up as a braiding factor, also called the monodromy $M$, that is a simple phase,  $M=e^{2i \pi \lambda}$. While in the latter, braiding is described by a monodromy factor which modulus is smaller than one, $|M|<1$ \cite{Lee22}. A quantitative understanding of interchannel interaction effects shall be important in this case, as non-abelian topological orders share with the Jain states their complex edge structure composed of several charge and neutral modes.

The experiments presented here can also be extended to the time or frequency domains. In the first case, one can implement the anyon version of the Hong-Ou-Mandel experiment \cite{Hong87,Bocquillon13}, where the emission of pulses carrying a fractional charge is trigerred. One then measures the variations of the current cross-correlations as a function of the time delay between the two sources \cite{Rech17,Jonckheere22}. In the second case, one generates a stationary flow of anyons, but measures the current cross-correlations at high frequency (instead of low frequency in the present experiment). It has already been shown that anyon properties could be directly measured from characteristic time or frequency scales, such as their fractional charge via the measurement of the Josephson frequency $f_J=e^*V/h$ \cite{Kapfer19,Bisognin19}. Implementing similar experimental methods in the anyon collider geometry would provide direct measurements of anyon fractional statistics \cite{Safi20,Jonckheere22}, with a simplified analysis that would not rely on the simultaneous analysis of the backscattering transmission $T$ of the central beam-splitter.

\section*{Acknowledgments}
We thank P. Degiovanni, B. Gr\'{e}maud, K. Iyer, T. Jonckheere, T. Martin, C. Mora, and B. Rosenow for useful discussions. This work is supported by the ANR grant "ANY-HALL", ANR-21-CE30-0064, and by the French RENATECH network.\\
\noindent\textit{Note added.} Our results are consistent with the independent investigation by P.~Glidic \textit{et al.} submitted jointly with the present work.  After writing this manuscript, we also became aware of the related work by J.Y.M.~Lee \textit{et al.} \cite{Lee22b} supporting braiding statistics at $\nu=1/3$ from auto-correlations in a single source setup (unbalanced case).

\section*{Appendix A: Non-linear $T(I_+)$ dependence}

In this appendix, we provide some details pertaining to the derivation of the central QPC transmission $T$ as a function of the input anyon current $I_+$. The derivation relies on a finite temperature generalization of Ref.~\onlinecite{Rosenow16}. In particular, the departure from zero temperature directly impacts the equilibrium correlation function of the tunneling operator $A$, which now reads
\begin{align}
\left\langle A (t) A^\dagger (0) \right\rangle_\text{eq} = \left| \zeta \right|^2   \left[ \frac{\sinh \left( i \pi \frac{k_B T_\text{el}}{\hbar} \tau_c \right)}{\sinh \left( \pi \frac{k_B T_\text{el}}{\hbar} \left( i \tau_c - t \right) \right)} \right]^{2 \delta} ,
\end{align}
where we recall that $\zeta$ is the tunneling amplitude at the QPC, $\tau_c$ is a short time cutoff and $\delta$ is the scaling dimension of the tunneling operator.

Substituting this back into the expression for the tunneling current at the central QPC, one has
\begin{align}
\left\langle I_T \right\rangle = 2 i e^* \left| \zeta \right|^2 \int dt & \left[ \frac{\sinh \left( i \pi \frac{k_B T_\text{el}}{\hbar} \tau_c \right)}{\sinh \left( \pi \frac{k_B T_\text{el}}{\hbar} \left( i \tau_c - t \right) \right)} \right]^{2 \delta} \nonumber \\
& \times  \frac{\sin \left( \frac{I_-}{e^*} t \sin 2 \pi \lambda  \right)}{\exp \left[ \frac{I_+}{e^*} |t| \left(1- \cos 2\pi \lambda \right) \right] } ,
\end{align}
which generalizes Eq.~(7) from Ref.~\onlinecite{Rosenow16} to the finite temperature case. At this stage, it is important to stress out that this expression still relies on the assumption that the source QPCs behave as Poissonian sources of anyons, which is only valid for temperatures $T_\text{el}$ such that $\frac{\hbar I_+}{2 \pi e^* k_B T_\text{el}} \gg 1$.

Introducing the new variables $u = \pi \frac{k_B T_\text{el}}{\hbar} t$ and $u_c = \pi \frac{k_B T_\text{el}}{\hbar} \tau_c$, this is further rewritten as
\begin{align}
\left\langle I_T \right\rangle =   \frac{2 \hbar e^* \left| \zeta \right|^2 }{\pi k_B T_\text{el}} &  \text{Re} \left\{ \int_0^\infty du \left[ \frac{\sinh \left( i u_c \right)}{\sinh \left( i u_c - u \right)} \right]^{2 \delta} \right. \nonumber \\
& \times e^{- 2 \xi_+ u} \left( e^{2 i \xi_- u} - e^{-2 i \xi_- u} \right) \Bigg\} ,
\end{align}
where we also defined the reduced variables $\xi_+ = \frac{\hbar I_+}{2 \pi e^* k_B T_\text{el}} \left(1- \cos 2\pi \lambda \right)$ and $\xi_- = \frac{\hbar I_-}{2\pi e^* k_B T_\text{el}}  \sin 2 \pi \lambda$.

The remaining integrals are readily obtained, as one can write
\begin{align}
\mathcal{I} (z) &= \int_0^\infty du  \left[ \frac{\sinh \left( i u_c \right)}{\sinh \left( i u_c - u \right)} \right]^{2 \delta} e^{2 i z u} \nonumber \\
&= \frac{1}{2} \left( 1 - e^{2 i u_c} \right)^{2 \delta} \int_0^\infty dx e^{(- \delta + i z) x} \left( 1 - e^{2 i u_c} e^{-x} \right)^{-2 \delta} \nonumber \\
&= \frac{\left( 1 - e^{2 i u_c} \right)^{2 \delta}}{2 \delta - 2i z} {}_2 F_1 \left( 2\delta, \delta-iz, 1+\delta-iz, e^{2 i u_c} \right) ,
\end{align}
where ${}_2 F_1$ is the Gauss hypergeometric function and we used known results to derive the final expression (see section 3.312.3 from Ref.~\onlinecite{Gradshteyn07}).

Keeping in mind that $u_c$ is vanishingly small, this can be further simplified, provided that $\delta < 1$ (and $\delta \neq 1/2$), as
\begin{align}
\mathcal{I} (z) &= \frac{1}{2} e^{- i \pi \delta} \left( 2 u_c \right)^{2 \delta} \frac{\Gamma \left( \delta-iz \right) \Gamma \left( 1 - 2 \delta\right)}{\Gamma \left( 1 - \delta - iz\right) } + \frac{i u_c}{1-2 \delta} ,
\end{align}
so that the tunneling current ultimately reads
\begin{align}
\left\langle I_T \right\rangle =  -  \frac{2 \hbar e^* \left| \zeta \right|^2 }{\pi k_B T_\text{el}}  &  \left( 2 u_c \right)^{2 \delta}  \Gamma \left( 1 - 2 \delta\right) \sin \pi\delta \nonumber \\
& \times \text{Im} \left[ \frac{\Gamma \left( \delta +i\xi_-  + \xi_+ \right) }{\Gamma \left( 1 - \delta +i\xi_-  + \xi_+ \right) } \right] .
\end{align}

The variations of the tunneling current with respect to the current difference $I_-$ at the input then reads, for the case of the balanced collider
\begin{align}
\left. \frac{\partial \left\langle I_T \right\rangle}{\partial I_-} \right|_{I_- = 0} &=    \left| \frac{\hbar \zeta }{\pi k_B T_\text{el}} \right|^2   \left( 2 u_c \right)^{2 \delta}  \Gamma \left( 1 - 2 \delta\right) \sin \pi\delta  \sin 2 \pi \lambda \nonumber \\
& \times  \frac{\Gamma \left( \delta + \xi_+ \right) }{\Gamma \left( 1 - \delta  + \xi_+ \right) }
\left[ \Psi \left(1 - \delta + \xi_+ \right) - \Psi \left( \delta + \xi_+\right) \right] ,
\end{align}
so that the transmission of the central QPC as a function of the input anyon current reads
\begin{align}
T = &  \alpha   \frac{\Gamma \left( \delta + \xi_+ \right) }{\Gamma \left( 1 - \delta  + \xi_+ \right) } \nonumber \\
& \times
\left[ \Psi \left(1 - \delta + \xi_+ \right) - \Psi \left( \delta + \xi_+\right) \right] ,
\end{align}
where $\alpha$ is a constant that does not depend on $I_+$.

\end{document}